\documentclass[reprint,superscriptaddress,amsmath,amssymb,aps,pra,floatfix]{revtex4-1}

\usepackage{graphicx}
\usepackage[usenames,dvipsnames]{xcolor}

\usepackage{hyperref}
\hypersetup{
  colorlinks,
  allcolors=NavyBlue,
  linktoc=all
}

\newcommand*\dagg{^{\dagger}}
\newcommand*{\unit}[1]{\ensuremath{\,\mathrm{#1}}}
\newcommand*{\s}[1]{\ensuremath{_\mathrm{#1}}}	
\newcommand*\C{\mathcal C}

\newcommand*\da{\delta\hat a}
\newcommand*\db{\delta\hat b}

\begin{document}

\title{Two-tone Optomechanical Instability and Its Fundamental Implications for Backaction-Evading Measurements}

\author{Itay Shomroni}\thanks{These authors contributed equally.}
\author{Amir Youssefi}\thanks{These authors contributed equally.}
\author{Nick Sauerwein}\thanks{These authors contributed equally.}
\author{Liu Qiu}\thanks{These authors contributed equally.}
\affiliation{Institute of Physics, \'Ecole Polytechnique F\'ed\'erale de Lausanne, CH-1015 Lausanne, Switzerland}
\author{Paul Seidler}
\affiliation{IBM Research--Zurich, S\"aumerstrasse 4, CH-8803 R\"uschlikon, Switzerland}
\author{Daniel Malz}
\affiliation{Max-Planck-Institut f\"ur Quantenoptik, Hans-Kopfermann-Stra{\ss}e 1, 85741 Garching, Germany}
\author{Andreas Nunnenkamp}
\affiliation{Cavendish Laboratory, University of Cambridge, Cambridge CB3 0HE, United Kingdom}
\author{Tobias J. Kippenberg}
\email{tobias.kippenberg@epfl.ch}
\affiliation{Institute of Physics, \'Ecole Polytechnique F\'ed\'erale de Lausanne, CH-1015 Lausanne, Switzerland}

\date{January 16th, 2019}

\begin{abstract}
While quantum mechanics imposes a fundamental limit on the precision of interferometric measurements of mechanical motion due to measurement backaction, the nonlinear nature of the coupling also leads to parametric instabilities that place practical limits on the sensitivity by limiting the power in the interferometer.
Such instabilities have been extensively studied in the context of gravitational wave detectors, and their presence has recently been reported in Advanced LIGO.
Here, we observe experimentally and describe theoretically a new type of optomechanical instability that arises in two-tone backaction-evading (BAE) measurements, a protocol
designed to overcome the standard quantum limit.
We demonstrate the effect in the optical domain with a photonic crystal nanobeam cavity and in the microwave domain with a micromechanical oscillator coupled to a microwave resonator.
In contrast to the well-known parametric oscillatory instability that occurs in single-tone, blue-detuned pumping, and results from a two-mode squeezing interaction between the optical and mechanical modes,the parametric instability in balanced two-tone optomechanics 
results from single-mode squeezing of the mechanical mode in the presence
of small detuning errors in the two pump frequencies.
Counterintuitively, the instability occurs even in the presence of perfectly balanced intracavity fields and can occur for both signs of detuning errors.
We find excellent quantitative agreement with our theoretical predictions.
Since the constraints on tuning accuracy become stricter with increasing probe power, the instability imposes a fundamental limitation on BAE measurements as well as other two-tone schemes, such as dissipative squeezing of optical and microwave fields or of mechanical motion.
In addition to identifying a new limitation in two-tone BAE measurements, the results also introduce a new type of nonlinear dynamics in cavity optomechanics.
\end{abstract}

\pacs{Valid PACS appear here}

\maketitle

\section{Introduction}

Interferometric position measurement of mechanical oscillators is the underlying principle of the Laser Interferometer Gravitational Observatory (LIGO)~\cite{ligo2016}
and constitutes one of the most sensitive techniques for determining absolute distance available to date.
In a similar vein, cavity optomechanical systems~\cite{aspelmeyer2014}, which exploit radiation-pressure coupling of light and mechanical motion in micromechanical and nanomechanical systems, have achieved some of the most sensitive measurements of mechanical motion relative to the zero-point motion~\cite{wilson2015,rossi2018}.
They can operate in a regime where measurement quantum backaction is relevant~\cite{purdy2013,teufel2016} and where cooling ~\cite{schliesser2008,chan2011,teufel2011} and amplification~\cite{braginsky2001,kippenberg2005,marquardt2006} via radiation-pressure backaction is accessible.
In both settings, the quantum fluctuations of radiation pressure place a fundamental limitation on the displacement sensitivity~\cite{caves1980,pace1993}.
Still, there can be other constraints.
Radiation-pressure nonlinearities
can equally well pose a limit to sensitivity.
Indeed, the parametric oscillatory instability~\cite{braginsky1967, braginsky1977, aguirregabiria1987, bel1988, braginsky2001,kippenberg2005,marquardt2006,ludwig2008}
is one of the most fundamental optomechanical effects
predicted to limit the performance of the LIGO detector by constraining the optical power below the self-induced oscillation threshold~\cite{pai2000,braginsky2001,braginsky2002,schediwy2004,ju2006,ju2006b,ju2006c,gurkovsky2007}.
It arises from the fact that
radiation-pressure coupling is intrinsically nonlinear, giving rise---in addition to static optical bistability~\cite{dorsel1983}---to rich nonlinear dynamics, leading to an intricate landscape of multiple stable attractors (dynamical multistability)~\cite{marquardt2006, ludwig2008, krause2015} and classical chaos~\cite{bakemeier2015}.
Radiation-pressure-induced parametric oscillatory instability has been observed in cavity optomechanical systems~\cite{kippenberg2005} and, a decade later, in the Advanced LIGO detector itself~\cite{evans2015}.
Special measures must be taken for its suppression~\cite{miller2011}.
Understanding such dynamical instabilities in optomechanical systems is important for the realization of ultrasensitive displacement measurements that operate with high cooperativity.
In addition, nonlinear phenomena in optomechanical systems have been the subject of experimental studies by themselves~\cite{krause2015,monifi2016,navarro-urrios2017}.

Here, we report 
a new type of instability caused by radiation pressure that is distinct from the parametric oscillatory instability.
The instability occurs in optomechanical systems driven with two tones.
One particular example is a class of quantum nondemolition measurements---two-tone backaction-evading (BAE) measurements as first proposed by Braginsky, Thorne, \textit{et al.}~\cite{thorne1978,braginsky1980,caves1980b}---that aim to surpass the standard quantum limit (SQL) of measurement of mechanical motion~\cite{caves1980b}.
These BAE measurements proceed by pumping an optomechanical system (that resides in the resolved sideband limit~\cite{schliesser2008}) simultaneously on the upper and lower motional sidebands of the cavity,
and allow in principle arbitrary measurement sensitivity; by increasing the probing power, measurement imprecision is decreased, \emph{without} incurring additional measurement noise due to quantum backaction.
Although theoretically proposed several decades ago, only recently have advances in cavity optomechanics made it possible to operate under conditions dominated by quantum backaction and carry out BAE measurements with both  microwave~\cite{suh2014,lecocq2015,lei2016,ockeloen-korppi2016} and optical~\cite{shomroni2019} systems.
The \textit{two-tone instability} reported here arises from deviations from the ideal BAE configuration where there is a finite-frequency detuning error with respect to both optical and mechanical resonance frequencies.
In contrast to the well-known parametric oscillatory instability that occurs in single-tone pumping on the upper motional sideband (or blue detuned in the bad-cavity limit), 
which results from a two-mode squeezing interaction (nondegenerate parametric down-conversion) between the optical and mechanical modes, and is associated with antidamping, the instability in balanced two-tone optomechanics
results from single-mode squeezing (degenerate parametric down-conversion) of the mechanical mode, and is associated with the optical spring effect.
While the parametric oscillatory instability is dynamically classified as an unstable spiral, the two-tone instability is classified as a saddle point~\cite{strogatz2015}.
We gain further insight by showing that the balanced two-tone scheme in the good-cavity (i.e. resolved-sideband) limit can be mapped to single-tone optomechanics in the bad-cavity limit.

The threshold for the onset of the instability depends on the magnitude of the tuning errors, and is also inversely proportional to the optical pump power.
For any given experimental inaccuracy in the pump frequency, a finite instability threshold
exists in two-tone experiments, which ultimately limits the maximum probe power and thus the achievable sensitivity.
As we show, these limitations can be prohibitive for strong pumping powers aiming to surpass the SQL.
We emphasize that the two-tone instability is intrinsic to the optomechanical interaction and does not arise from extraneous effects.
It depends exclusively on tuning errors and power.
Thus it stands in contrast to previously reported instabilities in BAE measurements associated with parametric driving~\cite{hertzberg2010,steinke2013}, where the underlying cause has been attributed to the dependence of the mechanical frequency on temperature~\cite{suh2012} or the presence of two-level systems~\cite{suh2013}.
In this sense, the two-tone instability poses fundamental constraints, and one may need to resort to active feedback techniques, as in the case of the parametric oscillatory instability~\cite{miller2011,harris2012}.

While our focus is on BAE measurements, it is important to note that the phenomenon reported here can affect other two-tone optomechanical protocols, such as dissipative optomechanical squeezing of optical and microwave fields ~\cite{kronwald2014,ockeloen-korppi2017} or of mechanical motion~\cite{kronwald2013,wollman2015,lecocq2015,pirkkalainen2015,lei2016}.
For example, in recent work on noiseless single-quadrature amplification of mechanical motion~\mbox{\cite{delaney2019}}, the squeezing effect we report here produces significant deviations from the expected system behavior.

\section{Observation of instability in two-tone pumping}
\label{sec:observation}

\begin{figure}
	\includegraphics[scale=1]{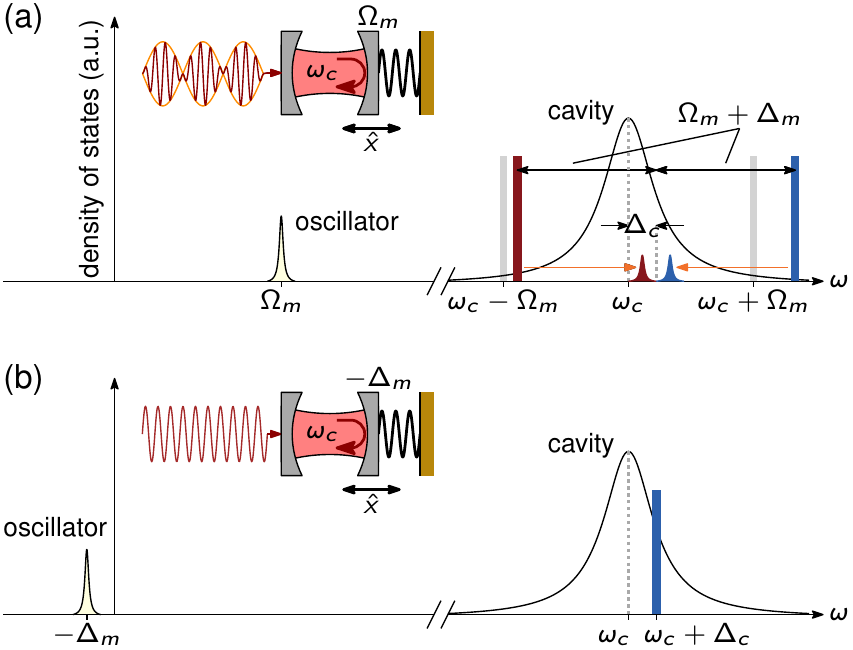}
	\caption{\textbf{Pumping scheme leading to two-tone instability.}
		(a)~Frequency-space representation of optical backaction-evading (BAE) measurement using two-tone pumping. An optomechanical system is pumped with two pumps that are placed on the lower and upper motional sideband of the cavity (shown in gray). Two detuning errors are introduced, 
		due to imperfect knowledge of the mechanical oscillator frequency ($\Delta_m$), and due to imperfect symmetric spacing around the cavities' resonance frequency $\omega_c$, as expressed by  $\Delta_c$.
		Also shown are the mechanical resonance at frequency $\Omega_m$ and the scattered mechanical sidebands.
		The inset shows an optomechanical system: a mechanical oscillator (position coordinate $\hat x$) that is the moving mirror of a Fabry-Perot cavity and that is coupled to the cavity mode by radiation pressure. 
		(b)~An equivalent system, in which the two-tone pumping is mapped to a Hamiltonian that exhibits the same dynamics as (a), consisting of a single continuous pump field applied at $\omega_c+\Delta_c$ and with the mechanical oscillator frequency obeying the substitution $\Omega_m\rightarrow -\Delta_m$.
		Note that $\Delta_c$ and $\Delta_m$ have been exaggerated for clarity.
	}
	\label{fig:freqs}
\end{figure}

In a BAE measurement the cavity, with resonance frequency $\omega_c$, is probed with two pump tones of equal power, each tuned to the upper and lower motional sideband of the cavity, i.e., at~$\omega_c\pm\Omega_m$.
It can also be understood as a single pump tuned at $\omega_c$ with full amplitude modulation at the mechanical frequency $\Omega_m$.
In Fig.~\ref{fig:freqs}(a) we illustrate the scheme, and introduce a small detuning error $\Delta_c\ll\kappa$ of the symmetrically spaced tones with respect to the cavity resonance, as well as an error $\Delta_m\sim\Gamma_m$ in modulation frequency.
As a result,  the two tones are detuned by $\Delta_c\pm(\Omega_m + \Delta_m)$ from the cavity resonance.
The motion of the oscillator, due to thermal noise from the environment and quantum backaction by the pump fields, is imprinted as fluctuations on the fields reflected from the optomechanical system.
The corresponding output noise spectrum of the two probes exhibits two Lorentzians separated by $2\Delta_m$, also shown in Fig.~\ref{fig:freqs}(a).
When $\Delta_c=\Delta_m=0$, an ideal BAE measurement is performed.
The mechanical sidebands are superimposed on each other, but while thermal motion adds in quadrature, the quantum backaction noise is canceled from the measurement record~\cite{clerk2008,suh2014,shomroni2019}.
A hallmark of BAE measurements is witnessing, as $\Delta_m$ is varied from a finite value to zero, a total mechanical noise that is lower than the sum total of the two individual mechanical noise spectra.
The total evaded backaction, expressed in units of mechanical quanta, is equal to the optomechanical cooperativity $\C$, proportional to the probing power.

\begin{figure*}
	\includegraphics[width=\linewidth]{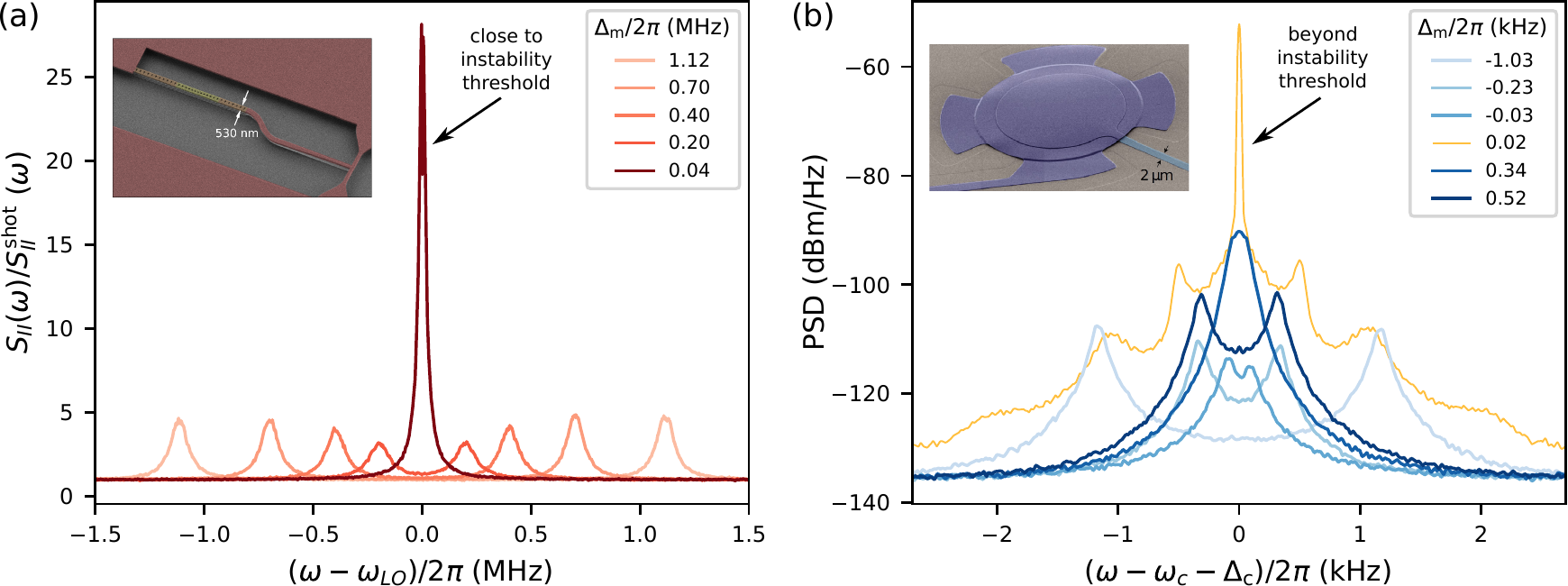}
	\caption{\textbf{Experimental observation of two-tone instability.}
		Panels~(a) and~(b) show two-tone BAE measurements in the presence of a small optical tuning error $\Delta_c$ and a mechanical tuning error $\Delta_m$.
		(a)~BAE measurement in the optical domain with an optomechanical system based on a silicon photonic crystal nanobeam cavity (inset).
		A sequence of measurements is shown, where the mechanical sidebands are measured via quantum-limited heterodyne detection, normalized to the shot-noise level.
		Here, $\omega\s{LO}$ is the optical frequency of the heterodyne local oscillator laser.
		In the sequence, the mechanical ``tuning error'' $\Delta_m$ is varied from a positive value toward zero, where $\Delta_m=0$ ideally corresponds to a BAE measurement.
		Because of the cavity tuning error $\Delta_c$, at $\Delta_m/2\pi=0.04\unit{MHz}$ a strong increase in the total mechanical noise is observed, as well as narrowing of the sideband, instead of the expected decrease due to backaction cancellation.
		This is the onset of the two-tone instability.
		(b)~BAE measurement in the microwave domain with a mechanically compliant capacitor coupled to a superconducting microwave resonator (inset).
		The procedure is the same as in (a).
		Measurements with both positive and negative values of $\Delta_m$ are shown.
		The measurement at $\Delta_m/2\pi=0.02\unit{kHz}$ occurs within the domain of instability (yellow curve), in which case the observed spectrum is distorted by saturation of the HEMT amplifier used in the detection chain.
		Note the logarithmic $y$ axis.
	}
	\label{fig:experimental}
\end{figure*}

Such BAE measurements have been carried out in the microwave domain in several experiments using mechanically compliant capacitors~\cite{suh2014,lecocq2015,lei2016,ockeloen-korppi2016},
and have recently been extended to the optical domain~\cite{shomroni2019}.
We carried out BAE experiments using an optomechanical photonic crystal nanobeam cavity operating at GHz frequencies mounted in a $^3$He buffer-gas cryostat, as detailed in prior work~\cite{qiu2018,shomroni2019}.
The parameters of the devices are shown in Appendix~\ref{sec:OptExperiment}, and details of the measurement setup, which uses a quantum-limited heterodyne detection method, are contained in Ref.~\onlinecite{shomroni2019}.
Figure~\ref{fig:experimental}(a) shows our BAE measurement in the optical domain.
In this experiment the cooperativity was set to $\C\simeq 2.6$ and $\Delta_m$ scanned from a positive value toward zero.
However, instead of the expected decrease in total noise, indicating BAE, an exponential increase in the total noise is observed near $\Delta_m/2\pi=40\unit{kHz}$.
Upon a further decrease in $\Delta_m$, the system leaves the linear regime as the optical cavity undergoes self-oscillations, i.e.,~we observe an instability.

To independently confirm the existence and universality of this phenomenon, we performed the measurement in an entirely different optomechanical system: an electromechanical system based on a mechanically-compliant vacuum-gap capacitor coupled to a superconducting microwave resonator placed in a dilution refrigerator~\cite{toth2017,bernier2017}.
Figure~\ref{fig:experimental}(b) shows this experiment, with a cooperativity of $\C=7$
(here, measurement backaction includes classical noise, which should also be canceled).
In this second measurement as well, we observe an exponential increase in the total noise as the detuning $\Delta_m$ is decreased.
As $\Delta_m$ is decreased further, the noise saturates the HEMT amplifier in the detection chain, leading to an increased noise floor [Fig.~\ref{fig:experimental}(b), yellow curve].
In both experiments, BAE is not observed under the given experimental conditions.

The origin of this instability is not a spurious effect in the experiments, as is evident from the observation that the same behavior occurs in two very different optomechanical systems measured with very different equipment.
Instead, as shown below, the instability is a direct consequence of the optomechanical interaction in the presence of the small tuning errors $\Delta_m$ and $\Delta_c$ and depends only on these two parameters and the cooperativity.
We next develop the theory behind this new instability in Sec.~\ref{sec:theory}, and perform a systematic experimental study in Sec.~\ref{sec:experiment} that fully confirms the theoretical predictions.

\section{Theory}
\label{sec:theory}

It is well known that the anti-damping induced by pumping the cavity on the upper motional sideband (or blue detuning in the bad-cavity limit) can induce a parametric oscillatory instability. 
In principle, there exists another type of dynamical instability in this system, 
one associated with the optical spring effect, i.e., a change in the restoring force induced by light.
This cannot occur in the resolved-sideband regime, in the relevant case of weak coupling between the mechanical mode and the cavity field.
Indeed, optomechanical systems typically employ high-quality-factor oscillators, where the shift in mechanical frequency due to dynamical backaction can be neglected.
However, as we show below, this instability may arise when pumping with \textit{two} tones
close to the upper and lower mechanical sidebands [Fig.~\ref{fig:freqs}(a)], as in BAE measurements, for example.
In fact, as we show, the situation when pumping with two tones and that for single-tone driving on the upper motional sideband are described by the same linearized equations.

We model the system by the standard optomechanical Hamiltonian comprising one cavity mode with frequency $\omega_c$, one mechanical mode with frequency $\Omega_m$, and a nonlinear interaction $\hat H\s{int}=-\hbar g_0\hat a\dagg\hat a(\hat b\dagg + \hat b)$, where $\hat a$~$(\hat b)$ denote the optical (mechanical) annihilation operator.
The cavity is driven by one or two coherent tones that produce a coherent intracavity field with amplitude $\bar a(t)$.
We move to the interaction picture with respect to the Hamiltonian
$\hat H_0 = \hbar\omega_l\hat a\dagg\hat a$
and linearize the operators, $\hat a\rightarrow \bar a(t)+\da$ and $\hat b\rightarrow\bar b+\db$, where $\bar a(t)$ is the coherent oscillation of the cavity field and $\bar{b}$ the static displacement of the mechanical oscillator.
This yields the linearized Hamiltonian
\begin{equation}
\hat H/\hbar = -\Delta_c\da\dagg\da + \Omega_m\db\dagg\db
 - g(t)(\da+\da\dagg)(\db+\db\dagg)
\end{equation}
where $g(t)=g_0\bar a(t)$ denotes the field-enhanced coupling.
We consider two distinct situations: single tone driving on the upper motional sideband and balanced two-tone driving on the upper and lower motional sidebands.
In single-tone driving, we have $\Delta_c=\Omega_m$ and $g(t)=g=\text{const}$, whereas the two-tone driving is described through $\Delta_c\approx 0$ and $g(t)=g\cos[(\Omega_m+\Delta_m)t]$.
Applying the rotating-wave approximation (RWA) in two-tone driving but not in single-tone driving, both situations can be described by the same Hamiltonian
\begin{equation}
  \hat H/\hbar=-\Delta_c\delta\hat a\dagg\delta\hat a-\Delta_m\delta\hat b\dagg\delta\hat b-g(\delta\hat a+\delta\hat a\dagg)(\delta\hat b+\delta\hat b\dagg),
  \label{eq:QND_hamiltonian}
\end{equation}
where in single-tone driving, $\Delta_m=-\Omega_m$, whereas in two-tone driving, $\Delta_m\approx0$.
This equivalence is illustrated in Fig.~\ref{fig:freqs}(b).

We describe the optomechanical system in terms of quantum Langevin equations~\cite{gardiner2004,aspelmeyer2014}, which take into account decay into the cavity and mechanical bath.
Eliminating the optical modes in frequency space, we arrive at an effective description for the mechanical mode only
\begin{equation}
  \left[ \frac{\Gamma_m}{2}-i(\Delta_m+\omega)+i\Sigma(\omega) \right]\delta\hat b(\omega)=-i\Sigma(\omega)\delta\hat b\dagg(\omega)+\hat\xi_{\mathrm{in}}(\omega),
  \label{eq:mech_solution}
\end{equation}
where the self-energy (effective coupling) is given through
\begin{equation}
  \Sigma(\omega)=\frac{2\Delta_c g^2}{(\kappa/2-i\omega)^2+\Delta_c^2}.
  \label{eq:eff_coupling}
\end{equation}
In Eq.~\eqref{eq:mech_solution} we have subsumed all noise contribution into a single generic noise input operator $\hat\xi_{\mathrm{in}}(\omega)$, which does not play a role in the instability mechanism.
So far we have not made any approximations (beyond the RWA in two-tone driving), and indeed Eq.~\eqref{eq:mech_solution} contains all the effects we wish to consider here. 
The self-energy $\Sigma$ plays two roles.
First, it couples $\delta\hat b\dagg$ to $\delta\hat b$, thus acting like the coupling in a degenerate parametric oscillator.
Second, as a self-energy, its real part renormalizes the frequency of the mechanical resonator and its imaginary part modifies the effective damping.

\emph{Parametric oscillatory instability.---}%
In single-tone driving on the upper motional sideband, a parametric oscillatory instability occurs if the optical antidamping overcomes the intrinsic damping. 
In this regime, the mechanical frequency is very large, so the right-hand side of Eq.~\eqref{eq:mech_solution} has hardly any effect and can be neglected in a RWA. 
The instability occurs because the mechanical damping is modified by the imaginary part of the susceptibility, Eq.~\eqref{eq:eff_coupling}, $-2\,\textrm{Im}[\Sigma(\Omega_m)]\simeq -\Gamma_{m}\C$,
where the cooperativity is $\C=4g^2/\kappa\Gamma_m$.
This yields the effective mechanical damping $\Gamma_{\mathrm{eff}}\simeq\Gamma_m(1-\C)$, thus
recovering the standard instability threshold $\C=1$.

\begin{figure}[t]
	\includegraphics[scale=1]{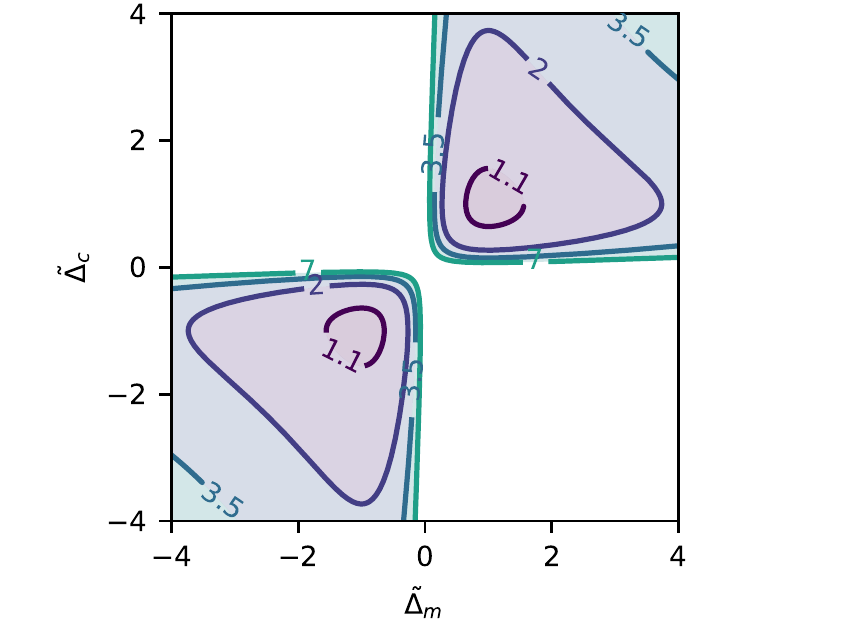}
	\caption{\textbf{Domains of two-tone instability.}
		The onset of the two-tone instability is given by the condition Eq.~\eqref{eq:condition} and depends only on 
		$\tilde\Delta_m\equiv\Delta_m/(\Gamma_m/2)$, $\tilde\Delta_c\equiv\Delta_c/(\kappa/2)$,
		and the cooperativity $\C$.
		Here we plot the domains of instability as a function of $\tilde\Delta_m$ and $\tilde\Delta_c$ for different cooperativities $\C$ (given as the contour labels).
		As $\C$ increases the stable region in the vicinity of the origin $\tilde\Delta_m=\tilde\Delta_c=0$ becomes smaller, reducing the range of $\tilde\Delta_m$ and $\tilde\Delta_c$ for which the system is stable.
	}
	\label{fig:contours}
\end{figure}

\emph{Two-tone instability.---}%
In backaction-evading measurements, however, $\Delta_m$ is small, as it represents a tuning error.
This makes the right-hand side of Eq.~\eqref{eq:mech_solution} near resonant, such that it cannot be neglected.
The instability arises in a way similar to a degenerate parametric oscillator~\cite{WallsMilburn2008}.
Since $\kappa$ is now a large parameter, we can neglect the frequency dependence of $\Sigma(\omega)\approx\Sigma(0)\equiv\Sigma\in\mathbb R$. Note that now the self-energy $\Sigma$ coincides with the optical spring effect due to a single drive detuned by $\Delta_c$ from cavity resonance.
This allows us to recast Eq.~\eqref{eq:mech_solution} again as an equation of motion
\begin{equation}
  \delta\dot{\hat b}(t)=\left[-\frac{\Gamma_m}{2}+i(\Delta_m-\Sigma)\right]\delta\hat b(t)-i\Sigma \delta\hat b\dagg(t)+\hat\xi_{\mathrm{in}}(t).
  \label{eq:eff_eom}
\end{equation}
This equation is the same quantum Langevin equation as one would write down for a damped degenerate parametric oscillator.
It can intuitively be viewed as arising from an optical spring effect modulated at $2(\Omega_m+\Delta_m)$.
The dynamical matrix corresponding to Eq.~\eqref{eq:eff_eom} has eigenvalues
$-\frac{\Gamma_m}{2}\pm i\sqrt{\Delta_m(\Delta_m-2\Sigma)}$.
For $2\Sigma<\Delta_m$, the eigenvalues have a negative real part (damping $\Gamma_m/2$) and a finite imaginary part (effective frequency $\Delta\s{eff}=\mathrm{Re}\,\sqrt{\Delta_m(\Delta_m-2\Sigma)}$).
As $\Sigma$ increases, first the effective frequency $\Delta_{\mathrm{eff}}$ vanishes, at which point the damping of the modes starts to be modified.
The vanishing of the effective mechanical frequency corresponds to the mechanical oscillation phase locking to the modulated optical field.
The threshold for instability occurs when
\begin{equation}
  4g^2\Delta_m\Delta_c=\left( \frac{\Gamma_m^2}{4}+\Delta_m^2 \right)\left( \frac{\kappa^2}{4}+\Delta_c^2 \right).
  \label{eq:condition}
\end{equation}
The instability threshold can also be written in terms of normalized detunings
$\tilde\Delta_c\equiv\Delta_c/(\kappa/2)$ and 
$\tilde\Delta_m\equiv\Delta_m/(\Gamma_m/2)$, yielding
$4\C=( 1+\tilde\Delta_c^2)( 1+\tilde\Delta_m^2)/(\tilde\Delta_c\tilde\Delta_m)$. 
This equation can only be fulfilled for $\C\geq1$, and we plot its contours in Fig.~\ref{fig:contours}.
Note that Eq.~\eqref{eq:condition} also predicts that the instability can occur for \emph{both} negative and positive values of the two detuning errors (provided they have the same sign), in contrast to the parametric oscillatory instability.
In Appendix~\ref{sec:eigenvalues} we derive instability regimes of both the parametric oscillatory and two-tone instabilities using the full Hamiltonian, Eq.~\eqref{eq:QND_hamiltonian}.

Apart from using locking techniques to reduce $\Delta_c$, it may be possible to hold off the onset of the two-tone instability by using active feedback.
Feedback techniques to counter the parametric oscillatory instability have been considered~\mbox{\cite{miller2011,harris2012}}.
In the case of the two-tone instability, the feedback force would naturally be applied on the measured quadrature (and as such is not of the viscous damping type).

\begin{figure*}[p]
\includegraphics[width=\linewidth]{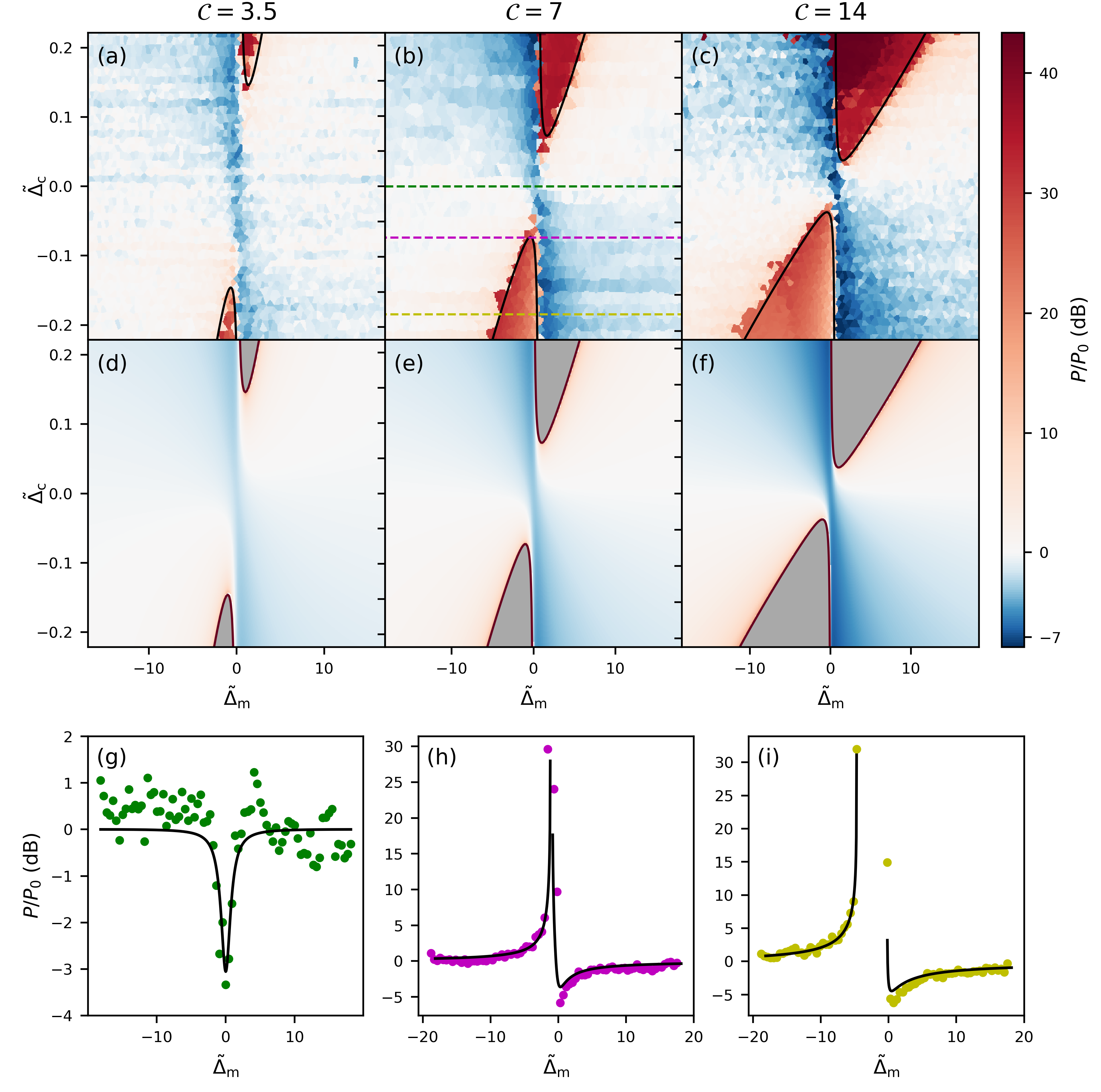}
\caption{\textbf{Investigation of the two-tone instability in a circuit-electromechanical system.}
(a)--(c) Mapping of the total mechanical noise as a function of $\tilde\Delta_m$ and $\tilde\Delta_c$, for cooperativities $\C=\{3.5,7,14\}$, respectively.
The total power $P$ in both mechanical sidebands in the output spectra is shown, normalized to the power $P_0$ at $(\tilde\Delta_m,\tilde\Delta_c)=(-18,0)$ (no tuning error and far from the BAE regime).
Since data points do not align on a regular grid (accounting for small changes in $\tilde\Delta_c$ along the horizontal scan), the rasterization was implemented using nearest-neighbor (Voronoi) partitioning.
Solid black lines are instability thresholds from Eq.~\eqref{eq:condition}.
(d)--(f) Theory plots corresponding to (a)--(c).
Gray areas are unstable regions predicted by theory.
(g)--(i) Cross sections of (b) for $\tilde\Delta_c=\{0,-0.07,-0.18\}$, respectively [nearest data points to horizontal dashed lines in~(b)].
Solid black lines are theory predictions based on the full linear model.}
\label{fig:MW}
\end{figure*}

\section{Experiment}
\label{sec:experiment}

To validate Eq.~\eqref{eq:condition}, the threshold for the two-tone instability in terms of $\Delta_m$ and $\Delta_c$, we perform a two-dimensional scan in the clean and well-controlled setting of circuit electromechanics using the system comprising a mechanically compliant vacuum-gap capacitor coupled to a superconducting resonant circuit placed in a dilution refrigerator~\cite{toth2017,bernier2017}.
Here, pump frequency fluctuations and cavity frequency fluctuations are significantly smaller than in the optical domain, and the detuning can be accurately controlled.
Full details on the system and experiment are given in Appendix~\ref{sec:MWexperimentDetails}.
We present measurements in the optical domain in Appendix~\ref{sec:OptExperiment}.

Figures~\ref{fig:MW}(a)--(c) show the total noise in the mechanical sidebands as a function of $\tilde\Delta_m$ and $\tilde\Delta_c$ for three different cooperativities $\C$.
Each horizontal cut in Figs.~\ref{fig:MW}(a)--(c) corresponds to a measurement of the type shown in Fig.~\ref{fig:experimental}(a).
The domains of instability are clearly evident as areas of increased noise (in red), in excellent agreement with Eq.~\eqref{eq:condition} (black contours).
In particular, instability only arises when $\tilde\Delta_m\tilde\Delta_c>0$, as predicted from Eq.~\eqref{eq:condition}, and can arise both for red- ($\tilde\Delta_c<0$) and blue-detuned ($\tilde\Delta_c>0$) mean probe frequency.
Figures~\ref{fig:MW}(d)--(f) show the theoretical plots corresponding to Fig.~\ref{fig:MW}(a)--(c), again in excellent agreement.
Figures~\ref{fig:MW}(g)--(i) show the horizontal cuts indicated in Fig.~\ref{fig:MW}(b).
The point $\tilde\Delta_m=\tilde\Delta_c=0$ corresponds to a ``perfect'' BAE measurement, as can be seen in Fig.~\ref{fig:MW}(g), where a $3\unit{dB}$ decrease in the total mechanical noise relative to $\tilde\Delta_m\neq 0$ due to cancellation of measurement backaction is evident.

In Appendix~\ref{sec:MWexperimentDetails} we show experimentally that, at the onset of instability, the effective mechanical frequency (due to the optical spring effect) equals the pump modulation frequency
(i.e.,~the effective mechanical frequency vanishes in the rotating frame).
In other words, while the instability occurs for $\Delta_m\neq 0$, in the experiment the sidebands coincide at the onset of instability.

The predicted decrease in size of the stable region with increasing pumping power is clearly evident in the experimental data depicted in Fig.~\ref{fig:MW},
with $\tilde\Delta_c\lesssim 1/2\C$ required to avoid instability
[e.g.,~$\tilde\Delta_c\lesssim 0.04$ for $\C=14$ in Fig.~\ref{fig:MW}(c)].
Overall, excellent agreement is obtained between theory and experiment, confirming our theoretical analysis and description of the effect.
Thus, optomechanics imposes strict tuning accuracy for a given measurement sensitivity in two-tone BAE measurements.
In this case, measurement backaction is due to both quantum and classical noise in the two microwave tones. It is important to emphasize that for BAE measurements that allow measurements beyond the SQL, cooperativities of $\mathcal{C} \gg \bar{n}_{m}$ are required, thus highlighting the stringent nature of the condition imposed by $\tilde\Delta_c\lesssim 1/2\C$.

\section{Conclusion}

We report experimentally and explain theoretically a new type of dynamical instability that was previously unreported in cavity optomechanics.
This instability is qualitatively different than the parametric oscillatory instability \cite{aguirregabiria1987,bel1988,fabre1994,pai2000,braginsky2001,marquardt2006,ludwig2008},
and originates from degenerate parametric amplification of the mechanical mode.
In the past, parametric oscillatory instability has limited certain single-tone experiments.  Our work now demonstrates that the performance of emerging optomechanical experiments, such as backaction-evading measurements aimed at surpassing the standard quantum limit~\mbox{\cite{suh2014,shomroni2019}}, generation of quantum squeezing~\mbox{\cite{kronwald2013,kronwald2014,woolley2013,woolley2014}}, and noiseless single-quadrature amplification~\mbox{\cite{delaney2019}}, will be intrinsically constrained by another instability determined by tuning accuracy and coupling strength.
Even below the instability threshold, these new dynamics need to be taken into account.

\emph{Data Availability.} 
The code and data used to produce the plots within this paper are available at \url{https://doi.org/10.5281/zenodo.3419929}.
All other data used in this study are available from the corresponding authors upon reason5able request.

\begin{acknowledgments}
This work was supported by funding from the Swiss National Science Foundation under Grant Agreement No. NCCR-QSIT: 51NF40-160591.
The samples were fabricated in the Center of MicroNanoTechnology (CMi) at EPFL.
The photonic sample was partially fabricated at the Binnig and Rohrer Nanotechnology Center (BRNC) at IBM Research--Zurich.
DM acknowledges support by the ERC Advanced Grant QUENOCOBA under the EU Horizon 2020 program (Grant Agreement No.~742102). 
A.N. acknowledges a University Research Fellowship from the Royal Society and support from the Winton Programme for the Physics of Sustainability.
This work was supported by the European Union's Horizon 2020 research and innovation programme under Grant Agreement No.~732894 (FET Proactive HOT).
\end{acknowledgments}

\appendix

\section{Eigenvalues of the dynamical matrix}
\label{sec:eigenvalues}

Further understanding of the two-tone instability, and its distinction from the parametric oscillatory instability, can be achieved by examining the eigenvalues of the dynamical matrix, similar to the analysis performed in Ref.~\cite{Malz2016}.
The Hamiltonian Eq.~\eqref{eq:QND_hamiltonian} leads to the quantum Langevin equations
\begin{subequations}
\label{eq:LangevinTwo}
\begin{align}
\label{eq:LangevinOpticalTwo}
\delta\dot{\hat a} &= -(\kappa/2-i\Delta_c)\da + ig(\db+\db\dagg) + \sqrt{\kappa}\da\s{in} \\
\delta\dot{\hat b} &= -(\Gamma_m/2-i\Delta_m)\db  + ig(\da+\da\dagg) + \sqrt{\Gamma_m}\db\s{in}.
\end{align}
\end{subequations}
Ignoring the input noise operators $\da\s{in}$ and $\db\s{in}$, which are irrelevant in the present analysis, the Langevin equations~\eqref{eq:LangevinTwo} can be written as a matrix equation $\dot{\mathbf{x}} = \mathbf{M}\mathbf{x}$, with
\begin{equation}
\label{eq:dynamicalmatrix}
\mathbf{M} = 
\left(
\begin{array}{cccc}
 -\kappa/2 & -\Delta_c  & 0 & 0 \\
 \Delta_c  & -\kappa/2 & 2 g & 0 \\
 0 & 0 & -\Gamma_m/2 & -\Delta_m  \\
 2 g & 0 & \Delta_m  & -\Gamma_m/2 \\
\end{array}
\right)
\end{equation}
and
\begin{equation}
\mathbf{x} = [\da+\da\dagg,\  i(\da\dagg-\da),\  \db+\db\dagg,\  i(\db\dagg-\db)]^T.
\end{equation}

We recall that this dynamical equation describes \textit{both} single-tone pumping, with the equivalence $-\Delta_m\rightarrow\Omega_m$, and two-tone pumping in the well-resolved sideband regime (Fig.~\ref{fig:freqs}).
An eigenvalue of the matrix Eq.~\eqref{eq:dynamicalmatrix} with a positive real part leads to an exponentially increasing solution, and thus signals an instability.

\begin{figure}
	\includegraphics[scale=1]{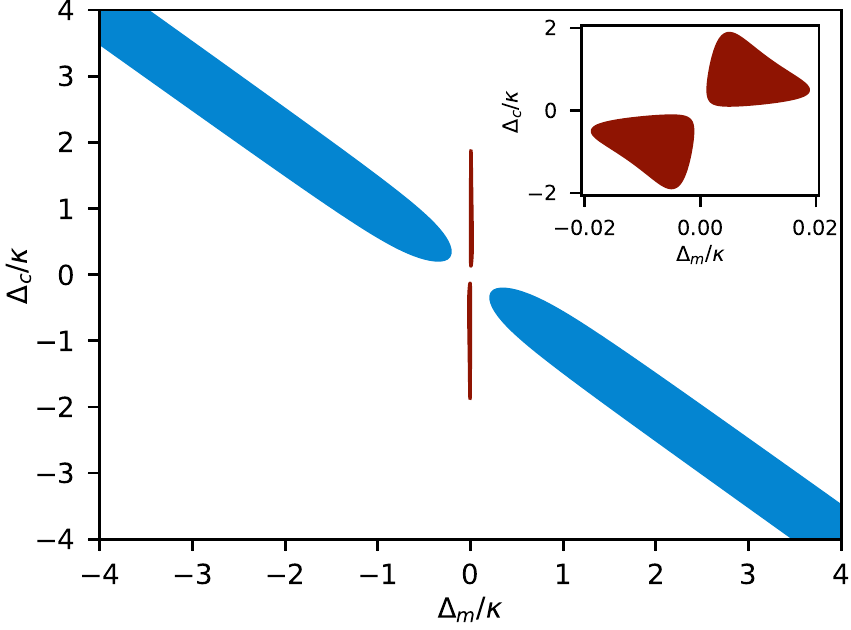}
	\caption{\textbf{Instability domains from eigenvalue analysis.}
		Domains where an eigenvalue of the dynamical matrix Eq.~\eqref{eq:dynamicalmatrix} has positive real part, signaling an instability.
		The blue domain, in the neighborhood of $-\Delta_m=\Omega_m\approx\Delta_c$, is the known parametric oscillatory instability.
		In this domain, the imaginary part of the eigenvalue is nonzero.
		In reality, the mechanical frequency is positive $\Omega_m>0$, such that the lower right-hand part is unphysical.
		The red domain that occurs for $\lvert\Delta_m\rvert\ll\kappa$ (enlargement shown in the inset) is the two-tone instability.
		In this domain $\Delta_m\Delta_c>0$, and the imaginary part of the eigenvalue is zero.
		The parameters used are $\kappa=1$, $\Gamma_m=10^{-2}$, and $\C=2$.
	}
	\label{fig:eigenvalues}
\end{figure}

Figure~\ref{fig:eigenvalues} shows the domains of instability in the parameter space spanned by $\Delta_m$ and $\Delta_c$. 
These domains separate into two classes, corresponding to the parametric and two-tone instabilities. In one class, which corresponds to the conventional parametric oscillatory instability, the imaginary part of the offending eigenvalue is nonzero, corresponding to spiral dynamics~\cite{strogatz2015}.
The onset of the parametric oscillatory instability coincides with the transition from a stable to an unstable spiral.
This class lies in the vicinity of the diagonal $\Delta_c\approx\Omega_m$, as expected (note that the regime $\Delta_m=-\Omega_m > 0$, although mathematically possible, is unphysical in this case).
The second class lies close to the origin, in particular $\lvert\Delta_m\rvert\ll\kappa$, and corresponds to the two-tone instability.
In this regime the eigenvalues are well approximated by the eigenvalues of Eq.~\eqref{eq:eff_eom} due to the slow dynamics of the optical field, and the instablity domains are given by the simple condition~\eqref{eq:condition}.
The eigenvalues are real and of opposite sign, corresponding to a saddle point, as expected from a degenerate parametric amplifier.
In this picture, as the power ($\propto\Sigma$) is increased, the dynamics change from a stable spiral to a stable node and finally to a saddle point~\cite{strogatz2015}.
Figure~\ref{fig:contours} shows the domains of two-tone instablity for different powers.
It is noteworthy to mention that Eq.~\eqref{eq:eff_eom} describes a harmonic oscillator with fixed damping $\Gamma_m$ and power-dependent natural frequency, i.e., $\ddot x+\Gamma_m\dot x+[\Gamma_m^2/4+\Delta_m(\Delta_m-2\Sigma)]x = F\s{ext}$. This system becomes unstable due to vanishing of the frequency (or restoring force).

\section{Microwave experiment details}
\label{sec:MWexperimentDetails}

For the microwave-domain part of this work, we used a system similar to the one described in Ref.~\onlinecite{bernier2017}.
Specifically, an overcoupled ($\kappa\s{ex}/2\pi = 2.65\unit{MHz}$, $\kappa_0/2\pi = 0.16\unit{MHz}$) Al superconducting microwave resonator with a resonance frequency of $\omega_c/2\pi=6.43\unit{GHz}$ that was coupled
(vacuum coupling rate $g_0/2\pi = 194\unit{Hz}$) to a mechanically compliant vacuum-gap capacitor ($\Omega_m/2\pi = 6.15\unit{MHz}$ and $\Gamma_m/2\pi \approx 20\unit{Hz}$).
The chip was cooled to about $15\unit{mK}$ in a dilution refrigerator.

In the experiment, we used three microwave sources with a common frequency reference to pump the optomechanical system:
two BAE pumps with frequencies $\omega_c+\Delta_c\pm(\Omega_m+\Delta_m)$ and an additional cooling pump  tuned to $\omega_c - \Omega_m-\delta\s{cool}$, where $\delta\s{cool}/2\pi = 400\unit{kHz}$.
The purpose of the cooling pump was 
to reduce the thermal occupation of the mechanical oscillator, thereby 
reducing the fluctuations in power of the mechanical sidebands 
that originate from fluctuations of the base temperature and vibrations of the dilution refrigerator,
since the oscillator is dominantly coupled to the microwave bath.
The cooling tone also increased the mechanical damping rate to $\Gamma\s{eff}/2\pi = 110\unit{Hz}$, allowing us to observe the narrowing of the mechanical sidebands when approaching the instability with better resolution.

To take the data in Fig.~\ref{fig:MW}, we first measured the cavity resonance frequency $\omega_c$ by applying the cooling tone alone and acquiring the microwave response \cite{NathanPhD}.
We then placed the BAE pumps symmetrically around the resonance, $\Delta_c=0$, and measured the mechanical resonance frequency $\Omega_m$ using the distance between the peaks of the mechanical sidebands.
The data of Fig.~\ref{fig:MW} were taken along horizontal scans, by first setting the required $\Delta_c$ and varying $\Delta_m$, acquiring noise spectra as in Fig.~\ref{fig:experimental}(b).
Prior to taking each horizontal line, we followed a procedure to ensure that no additional mechanical loss or amplification was introduced by the BAE pumps that might lead to parametric oscillatory instability:
First, only the red BAE pump was applied and its power optimized such that the width of the mechanical sideband corresponds to the desired single-tone cooperativity $\C$ (computed from the additional damping introduced).
Second, the blue BAE pump was added, and its power adjusted such that the width of the mechanical sideband narrows back to $\Gamma\s{eff}$.
This procedure was also followed prior to measurement of $\Omega_m$.

To account for cavity frequency fluctuations, the microwave frequency $\omega_c$ was also measured after each single point $(\Delta_m,\Delta_c)$.
This results in a slight vertical scatter along the horizontal scan.
We used nearest-neighbor (Voronoi) partitioning to present the full two-dimensional image.
In addition, we disconnected all pumps for $2\unit{s}$ between data points to let the mechanical oscillator thermalize and cancel any hysteresis effects due to the instability.

\begin{figure}
\includegraphics[scale=1]{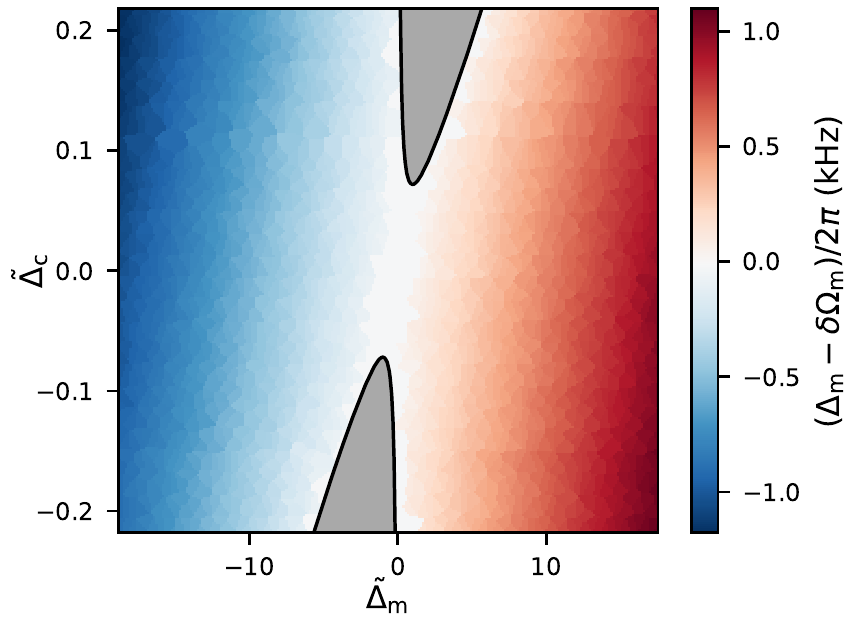}
\caption{\textbf{Vanishing of the effective mechanical frequency.}
The plot shows the difference between the measured effective mechanical frequency and pump modulation frequency as a function of the detunings.
Near the instability domain (shaded in gray), the two frequencies become equal, which corresponds to vanishing of the effective mechanical frequency in the rotating frame, confirming the theoretical treatment.
Measurement is the same as in Fig.~\ref{fig:MW}(b).
}	
\label{fig:fmechdev}
\end{figure}

\begin{figure*}[t]
	\includegraphics[scale=1]{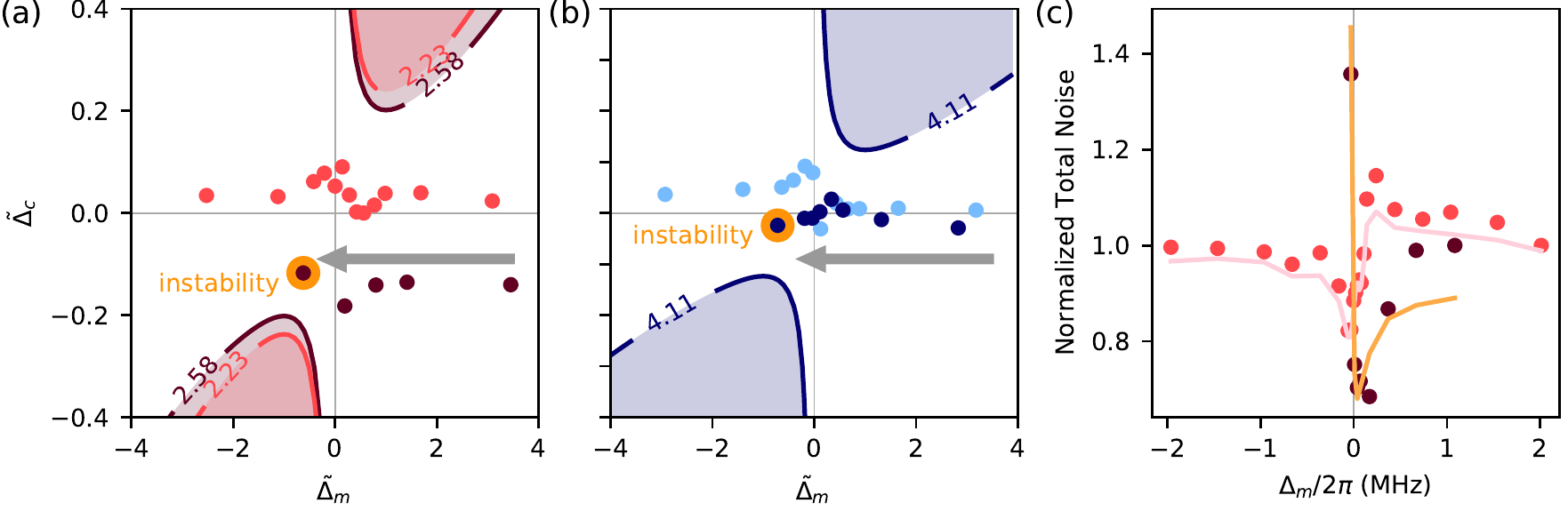}
	\caption{\textbf{Two-tone instability in the optical domain.}
		(a)~Two-tone BAE measurements (dark red and light red dots) scanning $\Delta_m$ with $\Delta_c$ kept approximately constant.
		The $(\tilde\Delta_m,\tilde\Delta_c)$ coordinate of each measurement is plotted.
		The instability contours for the respective cooperativities, around $\C\sim 2.2$--$2.6$ (contour label), are indicated with the same color.
		While one measurement sequence ($\C=2.23$) remains near $\tilde\Delta_c\approx 0$, the other ($\C=2.58$)  encounters instability in the vicinity of its corresponding contour.
		The highlighted dot is the measurement right before instability.
		(b)~Same as (a) with higher $\C\sim 4.1$.
		Here the stable region is smaller, highlighting the difficulty of achieving stable operation due to the inaccuracy in $\tilde\Delta_c$.
		(c)~Total noise power in the mechanical sidebands, normalized to 1 for large $\Delta_m$, for the two measurements sequences in (a).
		The sudden increase in power for the unstable data is evident.
		The solid lines are theoretical fits using the full Langevin equations.
		See text for more details.
	}
	\label{fig:optics}
\end{figure*}

In Fig.~\ref{fig:fmechdev} we plot the difference between $\Delta_m$ and the change in effective mechanical frequency of the oscillator $\delta\Omega_m$ (due to the optical spring effect), for the same data as in Fig.~\ref{fig:MW}(b).
Near the onset of the two-tone instability, the two sidebands coincide, and this difference approaches zero; i.e., the effective mechanical frequency $\Omega_m+\delta\Omega_m$ becomes equal to the modulation frequency of the pump, $\Omega_m+\Delta_m$.
This corresponds to vanishing of the effective mechanical frequency in the frame rotating with the modulation frequency, as predicted theoretically in Sec.~\ref{sec:theory}.

\section{Observation of two-tone instability in the optical domain}
\label{sec:OptExperiment}

As shown in Fig.~\ref{fig:experimental}(a), we also observed the two-tone instability in an optomechanical system operating in the optical domain.
The system is an optomechanical photonic crystal nanobeam cavity~\cite{eichenfield2009,chan2011} with a mechanical frequency $\Omega_m/2\pi\simeq 5.2\unit{GHz}$, optical resonance frequency $\omega_c\simeq 194\unit{THz}$ (wavelength $\lambda\simeq 1540\unit{nm}$), and cavity linewidth $\kappa/2\pi=300\unit{MHz}$ (optical $Q$ factor $\sim 6.5\times 10^5$).
The vacuum optomechanical coupling rate is $g_0/2\pi=930\unit{kHz}$.
The sample is measured in a $^3$He buffer-gas cryostat (Oxford Instruments HelioxTL), as reported previously~\cite{qiu2018,shomroni2019}.
The buffer gas facilitates the thermalization of the sample, preventing deleterious optical absorption heating and allows strong pumping.
The controlled pressure of the buffer gas affects the damping rate of the oscillator.
The various measurements reported here were done at temperatures in the range $4.6$--$4.9\unit{K}$, and pressures $32$--$160\unit{mbar}$, resulting in a damping rate of $\Gamma_m/2\pi=100$--$285\unit{kHz}$.

Figure~\ref{fig:optics} shows examples of the BAE measurements introduced in Sec.~\ref{sec:observation}, with $\Delta_m$ scanned from positive to negative values, while holding $\Delta_c$ approximately constant.
Two measurements of similar cooperativities $\C\sim 2.2$--$2.6$ are shown in Fig.~\ref{fig:optics}(a).
In the lower measurement $\Delta_c$ is farther from 0, bringing it in the vicinity of the domain of instability (see Fig.~\ref{fig:contours}), which is triggered on the next data point (not shown). 
This is the same measurement as Fig.~\ref{fig:experimental}(a).
The separation between the data and the domain of instability is $\sim 2\pi\times 15\unit{MHz}$, well within the uncertainty in our measurement of $\Delta_c$.
Figure~\ref{fig:optics}(b) shows similar data for higher cooperativity $\C\sim 4$, where uncertainty in measurement of $\Delta_c$ precludes discerning between the stable and unstable behavior.
Figure~\ref{fig:optics}(c) shows the total mechanical noise in the data of Fig.~\ref{fig:optics}(a), with the theoretical fit obtained from the Langevin equations~\eqref{eq:LangevinTwo}.
The data shown in light red, not encountering the instability, show imperfect, asymmetric BAE behavior (due to $\Delta_c\neq 0$).
The dark red data show the amplified noise prior to the onset of instability.

\clearpage

\bibliography{refs}

\end{document}